\documentclass[showpacs,aps,prb,preprint]{revtex4-1}
\usepackage{amsmath}
\usepackage{amsfonts}
\usepackage{mathrsfs}
\usepackage{bbold}
\usepackage{amssymb}
\usepackage{amsthm}
\usepackage{siunitx}
\usepackage{physics}
\usepackage{graphicx}

\newcommand{\sinc}{{\rm sinc}} 
\RequirePackage{color}
\usepackage{soul,xcolor}
\setstcolor{red}
\begin{document}
\title{Spin-orbit effects on the spin and pseudospin polarization in ac-driven silicene}
\author{Alexander L\'{o}pez}
\affiliation{Escuela Superior Polit\'ecnica del Litoral, ESPOL, Departamento de F\'isica, Campus Gustavo Galindo
 Km. 30.5 V\'ia Perimetral, P. O. Box 09-01-5863, Guayaquil, Ecuador}
 \email[To whom correspondence should be addressed. Electronic
 address: ]{alexander.lopez@physik.uni-regensburg.de}
 \author{Francisco Mireles}
\affiliation{Departamento de F\'isica Te\'orica, Centro de Nanociencias y Nanotecnolog\'ia, Universidad Nacional Aut\'onoma 
 de M\'exico, Apdo. Postal 14, 22800 Ensenada B.C., M\'exico}
 \author{John Schliemann}
 \affiliation{Institute of Theoretical Physics, University of Regensburg, Universit\"at Str. 31, 93053, Regensburg, Germany}
 \author{Benjamin Santos}
 \affiliation{INRS - \'Energie Mat\'eriaux T\'el\'ecommunications, Varennes, QC J3X 1S2, Canada}
\date{\today}
\begin{abstract}
We study the pseudospin and spin dynamical effects  in single-layer silicene due to a perpendicular electric field periodically driven and its interplay with the 
intrinsic and extrinsic (Rashba) spin-orbit interaction. We find that  the
spin nonconserving processes of the real spin of the quasiparticles in silicene, -- induced by the rather weak spin-orbit mechanisms --, manifest themselves as shifts of the 
resonances of its  quasienergy spectrum in the low coupling regime to the driving field. We show that there is an interesting cooperative 
effect among the, in principle, competing Rashba and intrinsic spin-orbit contributions. This is explicitly illustrated by exact and approximated analytical 
solutions of the dynamical equations. In addition, we show that a finite Rashba spin-orbit interaction is indeed necessary in order to achieve a
nonvanishing spin polarization. As additional feature, trivial and nontrivial topological phases might be distinguished from each other 
as fast or slow dynamical fluctuations of the spin polarization. We mention the possible experimental detection schemes of our theoretical results and their relevance 
in new practical implementation of periodically driven interactions in silicene physics and related two-dimensional systems.
\end{abstract}
\maketitle
\section{Introduction}
Silicene~\cite{takeda_theoretical_1994,
guzman-verri_electronic_2007,vogt_silicene:_2012,fleurence_experimental_2012,chen_evidence_2012}
is an interesting material belonging to the family of two-dimensional systems
that support massive Dirac fermions within their low energy regime. As well as 
graphene~\cite{novoselov_electric_2004,geim_rise_2007,castro_neto_electronic_2009},
its carbon counterpart, silicene has a honeycomb lattice structure with two interpenetrated
triangular sublattices.
Yet, in silicene, its lattice structure is naturally buckled. This in turn implies that 
fermions at each sublattice site will respond different to perpendicular
applied electric fields. Within the static scenario, previous works have shown
that applying a constant and perpendicular electric field, a topological phase transition
can be realized~\cite{ezawa_valley-polarized_2012}. In addition, recent proposals that
employ laser irradiation have shown to protect topological phases in Weyl
semimetals~\cite{gonzalez_topological_2017}. Studies of silicene in the nonequilibrium regime have  shown the realization of dynamical topological phases within the so-called off resonant approximation\cite{ezawa_photoinduced_2013, lopez_photoinduced_2015}.
Yet, these works neglected the Rashba spin-orbit interaction since it is
typically much smaller than the other energy scales in the problem.  We show in this work that
nevertheless, even a rather weak Rashba interaction in silicene can induce
a nontrivial spin polarization effect when a periodically varying electric field is
perpendicularly applied to the sample and as long as the system is in the resonant regime.

The study is focused within the Kane and Mele formulation for Dirac
fermions in the low energy approximation and investigate  the spin-pseudospin
effects due to a periodic time modulated perpendicular electric field
$E_\text{z}(t)$~\cite{kane_quantum_2005}. We show that a necessary requirement for
inducing a nontrivial dynamical evolution of physical observables, such as the
spin polarization, does require the nonvanishing of the Rashba spin-orbit interaction.
In this manner, our work differs from those previous approaches that have
overlooked the role of this spin nonconserving interaction in silicene. In order to tackle
the problem of silicene with periodically driven electric field, we make use of the
standard Floquet's theorem approach~\cite{shirley_solution_1965,grifoni_driven_1998,chu_beyond_2004}
and compare the exact numerical results with the rotating wave approximation (RWA) for
experimentally achievable parameter regimes.
We focus on the description of the physics close to resonance and near the critical value
of the static field that induces an energy gap closing of its spectrum. As it is
already well established in the literature, the behavior of the energy gap is important
in order to determine the topological phase of the driven system both in the static
regime~\cite{kane_z2_2005,hasan_colloquium_2010,bernevig_quantum_2006} as well as in the driven
scenario~\cite{kitagawa_topological_2010,lindner_floquet_2011,chen_high_2011,cayssol_floquet_2013,oka_photovoltaic_2009,zhou_optical_2011,calvo_tuning_2011}. We show that our results  of the 
nontrivial
spin dynamical effects are qualitatively dependent on the zero momentum bandgap.

The structure of the paper is as follows. In the next section we present the single particle
Dirac-like model for silicene under the presence of a time-varying electric field
and spin-orbit interaction and give numerical results for the quasienergy spectrum by
full diagonalization of the Floquet Hamiltonian. These exact numerical results are compared to
the approximate quasienergy spectra obtained via the RWA scheme for experimentally suitable
parameter regimes. Then, in section III we evaluate the spin and pseudospin
polarizations induced by the driving field using the analytical RWA solution
which allows us in turn to gain a better physical insight on the relevance of the resonantly
induced processes as well of the importance of the Rashba spin-orbit interaction.
In section IV we present concluding remarks and argue on the experimental
feasibility of the realization of our findings.
\section{Model}
We consider a monolayer silicene sample subject to a perpendicular static electric
field $E_\text{s}$. In addition, we consider a homogeneous time-dependent
electric field component $E_\text{z}(t)$ that drives the system out of equilibrium.
Then, following references~\cite{ezawa_valley-polarized_2012,ezawa_photoinduced_2013} the
long wavelength (low energy) approximation leads to the
$8\times8$ Dirac-like Hamiltonian ($\hbar=1$),
\begin{align}\label{e1}
\mathcal{H}_\eta(\vec{k},t)&=v_{\text{F}}(\eta k_x\sigma_x+k_y\sigma_y)+\eta s_z\sigma_z\lambda_\text{I}\\\nonumber 
&+\eta h_{11}\sigma_z+h_{22}-\ell [E_\text{s}+E_\text{z}(t)]\sigma_z,
\end{align}
where $v_\text{F}=\frac{\sqrt{3}at_b}{2}\approx \SI{e5}{\metre/\second}$ is the Fermi velocity for charge 
carriers in silicene with  $a = \SI{3.86}{\angstrom}$ the lattice constant
and $t_\text{b}=\SI{1.6}{\electronvolt}$ is the hopping parameter. Here
$\ell=\SI{0.23}{\angstrom}$ measures half the separation among the two sublattice planes.
In addition, $\eta=\pm 1$ describes the $K,K'$ Dirac points, $\sigma_i$ and $s_i$ are Pauli 
matrices describing pseudo and real spin degrees of freedom respectively, whereas $\vec{k}=\qty(k_x,k_y)$
is the in plane momentum measured from the Dirac point.
The parameter $\lambda_\text{I}=\SI{3.9}{\milli\electronvolt}$ represents the strength of the
effective intrinsic spin-orbit contribution within a tight binding approximation.
Furthermore, the two  contributions given by the terms $h_{11}= a\lambda_\text{R2}\qty(k_ys_x-k_xs_y)$ and 
$h_{22}= \lambda_{R1}\qty(\eta s_y\sigma_x-s_x\sigma_y)/2$ describe the Rashba spin-orbit coupling
associated to the nearest neighbor hopping and next-nearest neighbors, respectively.
It is found that $\lambda_{R2}=\SI{0.7}{\milli\electronvolt}$ and typically $h_{22}$ is much
smaller than the other energy scales in the problem~\cite{ezawa_valley-polarized_2012}.
Thus, we keep $h_{11}$ and  neglect $h_{22}$ in the following since our results will show that
this Rashba spin-orbit contribution is a necessary ingredient for nontrivial manipulation of the spin polarizaton of Dirac fermions.
Using the basis $\{\psi_{A\uparrow},\psi_{B\uparrow},\psi_{A\downarrow},\psi_{B\downarrow}\}$,
at the ${\bf K}$ Dirac point $\qty(\eta=+1)$ we have the reduced Hamiltonian
\begin{equation}\label{e2}
\mathcal{H}(\vec{k},t)=
\left(
\begin{array}{cccc}
E_{-}(t)&v_F k_-&iv_2k_-& 0 \\
v_F k_+&-E_{-}(t)&0& -iv_2k_- \\
-i v_2k_+&0&-E_+(t)& v_F k_-\\
0&iv_2k_+&v_F k_+& E_+(t) 
\end{array}
\right),
\end{equation}
where $k_\pm=k_x\pm ik_y$, $E_\pm(t)=\lambda_{I}\pm\ell\qty[E_s+E_z(t)]$ and $v_2=a\lambda_{R2}$. 
If we now define $\tan\phi=k_y/k_x$ and perform a unitary transformation with
$\mathcal{H}_1(k,t)=\mathcal{R}^\dagger_\phi\mathcal{H}(\vec{k},t)\mathcal{R}_\phi$ 
with $\mathcal{R}_\phi=\text{Diag}(e^{-i\phi},1,1,e^{i\phi})$ we get
\begin{equation}\label{e3}
\mathcal{H}_1(k,t)=
\left(
\begin{array}{cccc}
E_{-}(t)&v_F k&iv_2k& 0 \\
v_F k&-E_{-}(t)&0& -iv_2k \\
-i v_2k&0&-E_+(t)& v_F k\\
0&iv_2k&v_F k& E_+(t) 
\end{array}
\right).
\end{equation}
To simplify the notation we will denote $\lambda_{R2}=\lambda_R$ whenever we refer to the Rashba spin-orbit strength. Introducing the $4\times4$ matrix $\Lambda=s_x\otimes\mathbb{1}$, which explicitly reads
\begin{equation}\label{e0}
\Lambda=\left(
\begin{array}{cc}
 0& \mathbb{1}\\
\mathbb{1} & 0
\end{array}
\right),
\end{equation}
with $\mathbb{1}$ being the $2\times2$ identity matrix, the Hamiltonian takes a block diagonal form
\begin{equation}\label{t1}
 \mathcal{H}_2(t)=\mathcal{T}^\dagger_\alpha \mathcal{H}(t)\mathcal{T}_\alpha=
 \left(
\begin{array}{cc}
H_{2-}(t)& 0 \\
0&H_{2+}(t) 
\end{array}
\right),
\end{equation}
where the unitary transformation has the explicit form $\mathcal{T}_\alpha=\exp(-i\alpha\Lambda/2)$ and $\alpha$ is chosen to get rid of the off-diagonal terms. After some straightforward algebra one gets the condition for block diagonalization to fix the angle of transformation by the parameter relation 
$\tan\alpha=v_2k/\lambda_{I}$. Then, we find that
\begin{equation}\label{epm}
H_{2\pm}(t)=\mp\Delta_\pm\sigma_z+v_F k\sigma_x-\ell E_z(t)\sigma_z,
\end{equation}
where we have introduced the momentum-dependent static bandgap
$\Delta_\pm=\sqrt{\lambda_I^2+\qty(v_2k)^2}\pm V_z$,
with $V_z=\ell E_\text{s}$. In absence of the Rashba term $\qty(v_2=0)$, the static electric field has a critical value $E_s= E_c\equiv\lambda_{I}/\ell$ for which the static bandgap closes $\qty(\Delta_-=0)$
at one of the Dirac cones. This is the so-called single-valley Dirac cone phase. Thus, in the vicinity of the band inversion, one would expect that the typically small Rashba term proportional to $v_2$ might have a significant role in the dynamics, at least at large momenta.
This maximum value of the particle's momentum is, for consistency, determined by the 
validity of the linear dispersion approximation.
For instance, in a typical scattering problem this maximum value corresponds to the
Fermi momentum $k_\text{F}$. 
We are interested in determining the interplay among the Rashba nonconserving spin processes
and the modulation effects due to the driving field $\ell E_z(t)$, focusing in the near resonant
parameter regime where the RWA holds. Before proceeding further we perform an additional
unitary transformation $\mathcal{D}=\text{Diag}\{D_{-},D_+\}$, giving
\begin{equation}\label{t2}
 \mathcal{H}_3(t)=\mathcal{D}^\dagger \mathcal{H}_2(t)\mathcal{D}=
 \left(
\begin{array}{cc}
H_{3-}(t)& 0 \\
0&H_{3+}(t) 
\end{array}
\right),
\end{equation}
where $D_\pm=\sigma_z\cos\frac{\gamma_\pm}{2}\mp\sigma_x\sin\frac{\gamma_\pm}{2}$ and we have defined $\tan\gamma_\pm=v_Fk/\Delta_\pm$. We find then
\begin{equation}\label{hmenos1}
H_{3\pm}(t)=\mp\omega_\pm\sigma_z-\ell E_z(t)(\cos\gamma_\pm\sigma_z\mp\sin\gamma_\pm\sigma_x),
\end{equation}
where we have introduced the effective energy $\omega_\pm=\sqrt{\qty(v_Fk)^2+\Delta_\pm^2}$.
\noindent In order to explore the single-valley Dirac cone scenario beyond the static regime we consider a periodic time-dependent electric field 
$E_z(t+T)=E_z(t)$, where $T$ is the period.
Then, both of the Hamiltonians $H_{3\pm}(t)$ inherit its periodicity and we can resort to
Floquet theorem~\cite{shirley_solution_1965,grifoni_driven_1998,chu_beyond_2004} which
states that the general solution to the dynamics
\begin{equation}\label{t3}
i\partial_t\mathcal{U}_3(t)=\mathcal{H}_3(t)\mathcal{U}_3(t),
\end{equation}
can be written as
\begin{equation}\label{floquet}
\mathcal{U}_3(t)=\mathcal{P}(t)e^{-i \mathcal{H}_F t},
\end{equation}
with $\mathcal{P}(t)$ periodic and $\mathcal{H}_F$ a constant matrix, respectively.
The eigenvalues of $\mathcal{H}_F$ give the quasienergy spectrum of the periodically driven problem.
In order to determine these quasienergies one standard approach consists of
performing an expansion in the (infinite) eigenbasis of time periodic functions $F_n(t)=e^{in\Omega t}$,
where the frequency is defined as $\Omega=2\pi/T$. The resulting infinite eigenvalue problem is solved by a truncation procedure
to determine the Floquet exponents in such a way that adding more modes dot not qualitatively modify the obtained quasienergies.
In order to find the quasienergy spectrum one needs to evaluate the evolution
operator at $t=T$ and diagonalize the resulting matrix $\mathcal{U}_3\qty(T)$.
In the following section we find an approximate semi-analytical solution to the energy spectrum and compare to the exact result obtained via numerical diagonalization of the Floquet Hamiltonian in Fourier representation.
We also evaluate  the dynamics of the spin polarization and show that a finite Rashba spin-orbit interaction contribution is necessary in order to obtain nontrivial
dynamical effects on the spin polarization.
 \begin{figure}[t]
  \includegraphics[width=0.75\textwidth]{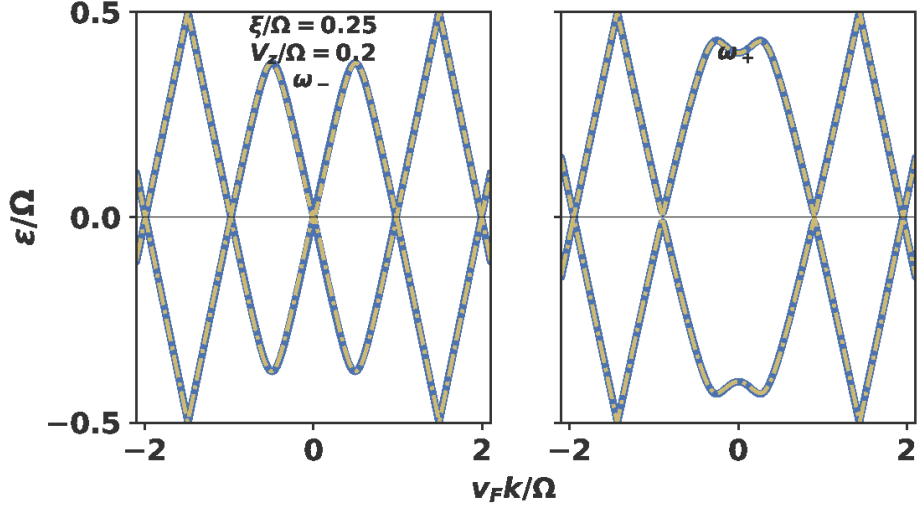} 
  \caption{\label{fig:1}(Color online)
  Exact (blue continuous line) and approximate (yellow dashed line) quasienergy spectrum for one of the
  effective spin components gapless (left panel)
  and the other gapped (right panel) at the Dirac cone $k_x=0$. This is the so-called single-valley Dirac cone
  configuration. The parameters have been chosen as
  $\xi=0.25\Omega$, $\lambda_{I}=V_z= 0.2\Omega$ and $\lambda_R=0.05\Omega$.}
  \end{figure}
\section{Analytical results within the RWA}
Let us now discuss an approximate solution leading to analytical expressions for the dynamical evolution of the physical quantities. This in turn
allows us to highlight the main physical mechanisms underlying the spin and pseudospin control of the Dirac fermions. 

\begin{figure}[t]
  \includegraphics[width=0.75\textwidth]{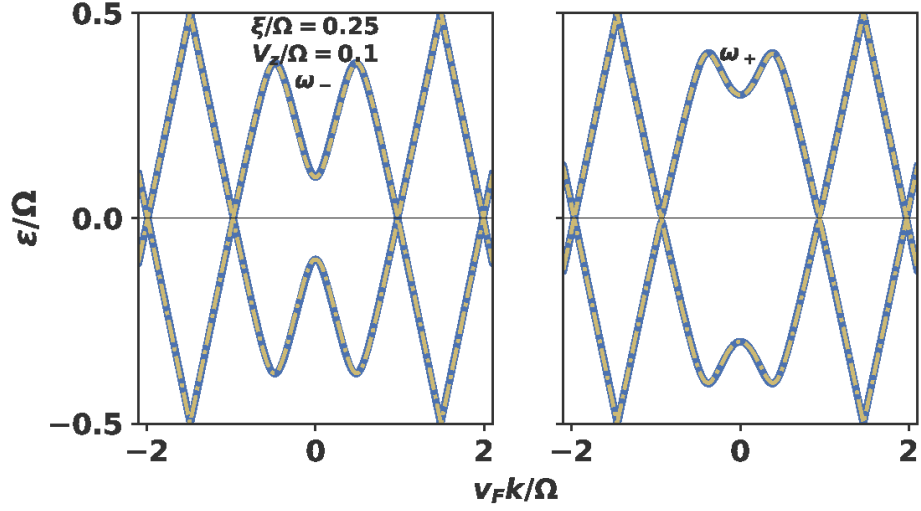} 
  \caption{\label{fig:2}(Color online) 
 Exact (blue continuous line) and approximate (yellow dashed line) quasienergy spectrum for the two effective
 spin sectors being gapless  at the Dirac cone $k=0$.
 The parameters have been chosen as $\xi=0.25\Omega$, $\lambda_{I}= 0.2\Omega$, $V_z=0.1\Omega$ and 
 $\lambda_R=0.05\Omega$.}
 \end{figure}

\noindent We begin by rewriting the upper and lower components of the block diagonal Hamiltonian
\begin{equation}\label{hmenos}
H_{3\pm}(t)=\mp\omega_\pm\sigma_z-\ell E_z\qty(t)\qty(\sigma_z\cos\gamma_\pm\mp\sigma_x\sin\gamma_\pm),
\end{equation}
where we recall that $\omega_\pm=\sqrt{\qty(v_Fk)^2+\Delta_\pm^2}$.
Changing to the interaction representation the dynamics is dictated by
$V_{I_\pm}\qty(t)=e^{\mp i\omega_\pm t\sigma_z}V_\pm\qty(t)e^{\pm i\omega_\pm t\sigma_z}$,
where $V_\pm\qty(t)$ is the time-dependent contribution shown in equation~(\ref{hmenos}).
Choosing the time dependence of the driving field as $\ell E(t)=\xi\sin\Omega t$ and using the algebra of 
the Pauli matrices we get,
\begin{align}\label{vi}
V_{I_\pm}(t)=&-\xi\sin{\Omega t}\left[\sigma_z\cos{\gamma_\pm}\mp
\sin{\gamma_\pm}\right.\nonumber\\
&\left. \qty(\cos{2\omega_- t}\sigma_x\pm\sin{2\omega_- t}\sigma_y)\right].
\end{align}
In order to implement the rotating wave approximation (RWA) we would like to remind that it is a standard 
approximation scheme that renders the Floquet Hamiltonian
time-independent by restricting to small values of the effective coupling strength to the driving field and for frequencies with values 
near those of 
the energies of the static Hamiltonian, i.e., near resonances. Hence, if we invoke the RWA, we get the approximate interaction contribution
\begin{align}\label{vrwa}
V_{I\pm}(t)\approx&\frac{\xi\sin{\gamma_\pm}}{2i}(e^{\mp i\delta_\pm t}\sigma_+- e^{\pm i\delta_\pm t}\sigma_-), 
\end{align}
where the frequency detuning that characterizes near resonance contributions is defined
as $\delta_\pm=2\omega_\pm-\Omega$. In addition, we have used the standard definitions
$\sigma_\pm=\qty(\sigma_x\pm i\sigma_y)/2$.
To get~(\ref{vrwa}) we have assumed $\Omega-2\omega_\pm\sim 0$ and thus have disregarded
the quickly oscillating terms proportional to $\pm\Omega$ as well as those corresponding
to the secular contributions $\Omega+2\omega_\pm$.
Switching back to the Schr\"{o}dinger representation we get the effective RWA Hamiltonian
\begin{equation}\label{hrwa}
h_{\pm}(t)=\mp\sigma_z\omega_\pm-i g_\pm\qty(e^{\pm i\Omega t}\sigma_+-e^{\mp i\Omega t}\sigma_-),
\end{equation}
where we have introduced the effective momentum-dependent coupling constant $g_\pm$, that reads explicitly
\begin{equation}
 g_\pm=\frac{\xi\sin\gamma_\pm}{2}=\frac{\xi v_Fk}{2\omega_\pm}.
\end{equation}
The Hamiltonian~(\ref{hrwa}) is taking to a time-independent form, $h_{F\pm}$, by means of periodic unitary transformations in each subblock matrix
$P_\pm\qty(t)=e^{\pm i(\mathbb{1}+\sigma_z)\Omega t/2}$, which relies on the standard transformation rule
$h_{F_\pm}=P_\pm ^\dagger(t)h_\pm(t)P_\pm (t)-i P_\pm ^\dagger(t)\partial_tP_\pm(t)$.
Explicitly, we have
\begin{equation}
 h_{F_\pm}=\pm\frac{\Omega}{2}\mathbb{1}\mp\frac{2\omega_\pm-\Omega}{2}\sigma_z+ g_\pm\sigma_y,
\end{equation}
for which the approximate quasienergies are found to be given as
\begin{align}
\varepsilon_{\nu\pm}=&\frac{1}{2}\qty(\nu\sqrt{4g_{\pm}^2+(2\omega_\pm-\Omega)^2}\pm\Omega),
\qquad\textrm{mod}\,\Omega\nonumber\\
\end{align}
where $\nu=\pm1$ is the energy sub-band index. Since the Hamiltonian preserves its
in-plane rotational invariance, we set from now on $k_y=0$, without loss of generality and denote $k_x=k$.
In~\figurename{~\ref{fig:1}} we show the quasienergy spectrum for the configuration when one spin component
is gapped  and the other gapless at $k=0$. The continuous line is the exact result obtained from
a numerical diagonalization of both $H_{3\pm}(t)$, whereas the dashed line
corresponds to the RWA scheme for a set of parameters which has been set as $\lambda_I=V_z=0.2\Omega$ and $\lambda_R=0.05\Omega$.
We find an excellent agreement between the RWA and exact results for an effective value of the coupling strength $\xi=0.25\Omega$.
Indeed, we have checked that the RWA properly describes the quasienergies even for values of the effective coupling strengh as large
as $\xi=0.5\Omega$ where they begin to differ from each other. If we choose the set of parameters $\lambda_I=0.2\Omega$ and $V_z=0.1\Omega$
with $\lambda_R=0.05\Omega$
same coupling strength $\xi=0.25\Omega$, we find that both pseudospin sectors are gapped at $k=0$, as it is shown in~\figurename{~\ref{fig:2}}.
\\
Having checked the quasienergy spectrum, we turn our attention to the interplay of the Rashba spin-orbit coupling and driving field. Previous works have 
neglected the Rashba spin-orbit in silicene\cite{ezawa_valley-polarized_2012, ezawa_photoinduced_2013, lopez_photoinduced_2015} because it possesses a value much smaller than the other energy scales in the problem.
Our purpose now is to show that it does have an important role in determining the time evolution of the spin polarization $S_z(t)$
and its interplay with the ac-driven component of the electric field renders nontrivial the time evolution of $S_z(t)$ and although small,
we show that $\lambda_R$ is indeed a necessary ingredient for getting a finite $S_z(t)$ for a properly chosen initially prepared state. 
Our purpose is to show how the dynamics of the Dirac fermions can be modified via this parameter. Let us discuss these issues in detail.
We first note that the block-diagonal structure of the Hamiltonian allows us to define an effective spin corresponding to these subblocks. Thus, we could expect that
this block diagonal condition could imply that effective spin is an approximate conserved quantity. Yet, in the following we show this is not the case due to the competing effects of
the Rashba spin-orbit interaction and the driving field. In order to quantify the 
robustness of spin polarization under the driving field and the spin nonconserving effects due to the Rashba 
spin-orbit coupling, we evaluate now the time evolution of the spin polarization $S_z(t)$, defined as
\begin{equation}
S_z(t)= \bra{\Psi(0)}\mathcal{U}^\dagger(t)\mathcal{S}_z\mathcal{U}(t)\ket{\Psi(0)},
\end{equation}
where $\ket{\Psi(0)}$ is the initially prepared electronic state and $\mathcal{S}_z=s_z\otimes \mathbb{1}$.
The explicit form of the evolution operator ${\cal U}(t)$ would be given below.
In order to highlight the role of the driving field and Rashba term we will choose the initial
state $\ket{\Psi(0)}$ as
\begin{equation}
 \ket{\Psi(0)} = \frac{1}{\sqrt{2}}\left(
 \begin{array}{c}
  \ket{\psi_-(0)}\\
  \ket{\psi_+(0)}
 \end{array}
\right),
\end{equation}
 \begin{figure}[t]
  \includegraphics[width=0.75\textwidth]{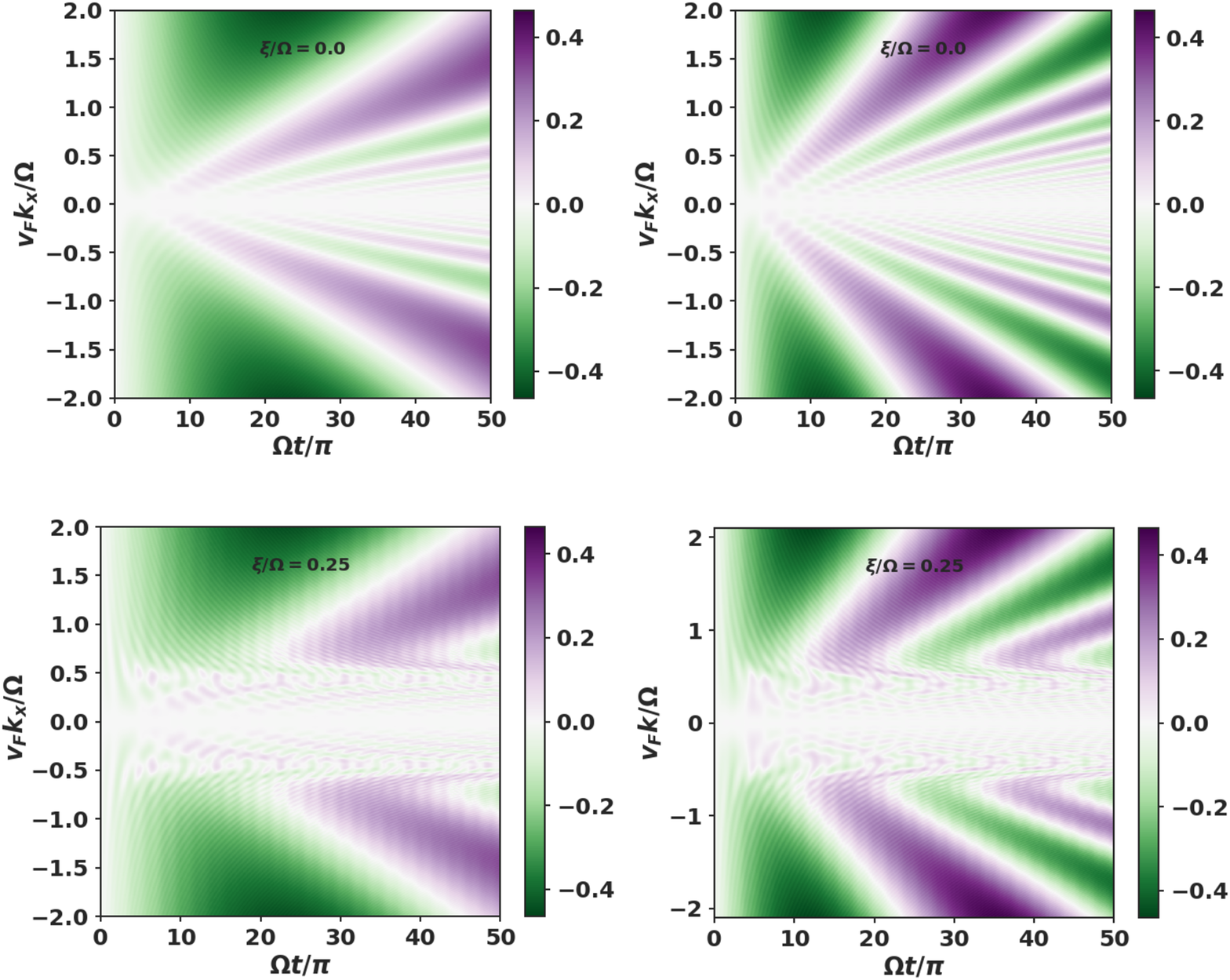}
  \caption{(Color online) 
   Effective spin polarization $S_z(t)$ as a function of time and momentum for $\lambda_R=0.05\Omega$.
   The upper panels correspond to vanishing $\xi=0$ with left (right) panel having the other parameters
   as given in~\figurename{~\ref{fig:1}} (\figurename{~\ref{fig:2}}), respectively.
   The lower left (right) panel shows the $S_z(t)$
   for  finite $\xi=0.25\Omega$  as in~\figurename{~\ref{fig:1}} (\figurename{~\ref{fig:2}}). We notice that
   near the Dirac point
   ($k=0$) additional spin inversion flips signal the competing effects among the Rashba and driving field.
   For large momenta the Rashba contribution dominates and we observe no dynamical effects due to the driving
   field.}
   \label{fig:3}
 \end{figure}
with vanishing initial spin polarization
$S_z(0)=\qty(\braket{\psi_-(0)}{\psi_-(0)}-\braket{\psi_+(0)}{\psi_+(0)})/2=0$. The simplest choice is to set
$s_z\ket{\psi_\pm(0)}=\pm\ket{\psi_\pm(0)}$. Using the transformations that render the evolution operator block diagonal, it reads explicitly
\begin{equation}
\mathcal{U}_3(t)=\left(
\begin{array}{cc}
U_{3-}(t)&0\\
0&U_{3+}(t)
\end{array}
\right)
\end{equation}
with
\begin{equation}\label{upara}
U_{3\pm}(t)=\left(
\begin{array}{cc}
a_\pm(t)&b_\pm\\
-b^{*}_\pm&a^{*}_\pm(t)
\end{array}
\right)
\end{equation}
where the dimensionless functions $a_\pm(t)$ and $b_\pm(t)$ read for the RWA result as
\begin{align}\label{rwaso}
a_\pm(t)=&e^{\pm i\Omega t/2}\qty[\cos \Gamma_\pm t\pm\frac{i\delta_\pm t}{2}\sinc \Gamma_\pm t]\\
b_\pm(t)=&g_\pm te^{\pm i\Omega t/2}\sinc \Gamma_\pm t,
\end{align}
and satisfy the restriction
\begin{equation}
\abs{a_\pm(t)}^2+\abs{b_\pm(t)}^2=1.
\end{equation}
In addition, we have defined $\Gamma_\pm=1/2\sqrt{4g_\pm^2+\delta^2_\pm}$,
with $\delta_\pm=2\omega_\pm-\Omega$ and we remind that $\tan\alpha=v_2k/\lambda_I$
measures the ratio of the two spin-orbit contributions that render the Hamiltonian block diagonal
as was described previously. Thus we get
\begin{align}\label{sz1}
S_z(t)&=\frac{ \sin\alpha}{2}\left\{ \textrm{Re}\qty[W_{--}(t)-W_{++}(t)]\sin\alpha\right.
\nonumber\\
&+\left. \textrm{Im}\qty[W_{-+}(t)-W_{+-}(t)]\cos\alpha+\textrm{Im}\qty[W_{+-}(t)+W_{+-}(t)]\right\},
\end{align}
where $W_{--}(t)=\bra{\psi_-(0)}W(t)\ket{\psi_-(0)}$, and so on, are the matrix elements of the operator
\begin{equation}\label{w}
 W(t)=D_-U^\dagger_{3-}(t)D_-D_+U_{3+}(t)D_+.
\end{equation}
After some algebraic operations we get the result,
\begin{align}\label{sz2}
S_z(t)
=&\sin\alpha\times\nonumber\\
&\left[\qty(\textrm{Im}\, a_-\textrm{Re}\, b_++\textrm{Re}\, a_+\textrm{Im}\, b_-)
\cos\gamma_--\right.\nonumber\\
&\qty(\textrm{Im} b_+\textrm{Re} a_-+\textrm{Im} a_+\textrm{Re} b_-)\cos\gamma_++\nonumber\\
&\qty(\textrm{Re}\, b_+\textrm{Im}\, b_--\textrm{Re}\, a_+\textrm{Im}\, a_-)\sin\gamma_-+\nonumber\\
&\left.\qty(\textrm{Im} b_+\textrm{Re} b_--\textrm{Im} a_+\textrm{Re} a_-)\sin\gamma_+\right],
\end{align}
Since the parameterization of the evolution operator given in equation~(\ref{upara}) is valid in general, we must remark that the result given in equation~(\ref{sz2})
is exact and thus allows us to verify both, the numerical calculation and the RWA result. Indeed, the functional form of the dimensionless functions~(\ref{rwaso}) for 
the exact result would differ from the relations corresponding to the RWA, but the polarization would still be calculated from expression~(\ref{sz2}), with the corresponding 
dimensionless quantities for the exact result. 
Since we have verified that RWA is valid for values of the coupling strength 
close to $\xi=0.5\Omega$ we rely on the physics within the RWA which is encoded
in the explicit form of the functions defined in equations~(\ref{rwaso}). Moreover, the result given 
in equation~(\ref{sz2}) shows that the vanishing of the first term in the parenthesis of equation~(\ref{sz1}) means that all polarization effects are first order in the Rashba spin-orbit strength
since $\sin\alpha\propto \lambda_R$
whereas the vanishing of the second term can be explained via the fact that it is proportional to the 
intrinsic spin-orbit interaction $\cos\alpha\propto\lambda_I$ which should not induce
any spin-flipping processes.
\\
Let us now discuss the results for the spin polarization $S_z(t)$ within the RWA as given in equation~(\ref{sz2}). In~\figurename{~\ref{fig:3}}
we show a density plot of $S_z(t)$ as a function of momentum and time for different parameter configurations. The upper panels show the dynamical evolution
of the polarization in absence of the driving field $\qty(\xi=0)$ for the single-valley Dirac cone
configuration~\figurename{~\ref{fig:1}} and gapless (\figurename{~\ref{fig:2}}) scenario.
Yet, as seen in the two lower panels of~\figurename{~\ref{fig:3}}, once $\xi=0.25\Omega$ is finite a richer physical scenario is observed.
In general, it is observed that spin-flipping processeses only require a finite value of the Rashba spin-orbit interaction and thus it should be considered as a valuable
parameter in experimental setups. We would like to remark that previous
works~\cite{ezawa_valley-polarized_2012,ezawa_photoinduced_2013, lopez_photoinduced_2015} have disregarded the role of the Rashba spin-orbit coupling in the physics of
monolayer silicene based on its rather small value as compared to the other energy scales in the problem.
However, our results for the dynamical evolution of the spin polarization show that for vanishing
$\lambda_R$, there are not pseudospin-spin
oscillations and $S_z(t)$ vanishes exactly for all values of the time and momentum variables for the initially chosen state. As discussed before, mathematically speaking, this can be seen by the
$\sin\alpha$ prefactor in equation~(\ref{sz2}), but the physical reason for this result relies essentially in the nonconserving nature of the
Rashba coupling. In presence of the driving field $\qty(\xi=0.25\Omega)$,
the dynamics of $S_z(t)$ is determined by the effective spin sectors being in the 
configuration of~\figurename{~\ref{fig:1}} (lower left panel) or~\figurename{~\ref{fig:2}} (lower right panel) of the quasienergies shown with $V_z=\lambda_I$ $\qty(V_z\neq\lambda_I)$. In the first situation, the dynamics
could be said to be {\it fast}, meaning that, within the same time window, more spin-flipping occurs for those
quasienergies corresponding to the so-called single-valley Dirac cone configuration of~\figurename{~\ref{fig:1}}\cite{ezawa_photoinduced_2013}. In this manner,
the analysis of the dynamics of $S_z(t)$ could afford an experimental verification for the topological features of Dirac fermions in silicene and related materials. Moreover, for momenta in the 
vicinity of the Dirac point $k=0$, the role of the driving field is enhanced since it qualitatively modifies the time evolution of $S_z(t)$ which can in principle be understood
as the resonant processes that lead to additional spin-flips and steems from the noncommuting nature of the Rashba spin-orbit interaction and the driving electric 
field and amounts to a direct manipulation of the pseudospin degree of freedom as we show below.
\\
Therefore, we find that even though the single-valley Dirac cone configuration $E_z=E_c$ might be expected as the only
requirement for a nontrivial spin-pseudospin behavior of the Dirac fermions on a single layer of silicene, this
is not the only parameter condition that renders nontrivial behavior on other quantities of interest, such as
the spin polarization. Thus, our results show that indeed the, in principle, competing effects of both the
Rashba term and the oscillating electric field are necessary in order to get a nontrivial evolution of the spin polarization. Of course, the Rashba term encodes the pseudospin-spin coupling and accounts for the angular
momentum exchange among these two degrees of freedom whereas the resonant nature of the driven field is
encoded in the $\Omega$ condition as well and the $\xi$ coupling strength that render the RWA valid.
\\
In order to further explore the dynamical features of the model, we now focus on the pseudospin polarization dynamics, given as
  \begin{equation}
  P_z(t)=  \bra{\Psi(0)}\mathcal{U}^\dagger(t)\mathcal{P}_z\mathcal{U}(t)\ket{\Psi(0)},
  \end{equation}
with $\mathcal{P}_z=\mathbb{1}\otimes\sigma_z$. After some algebraic manipulations we get the result
\begin{align}\label{pz}
P_z(t)
=&\cos\alpha\times\nonumber\\
&\left[1/2\qty(\cos^2\gamma_--\sin^2\gamma_-)
   \qty(\textrm{Im}~a^2_--\textrm{Im}~b^2_-)
   +1/2\qty(\textrm{Re}~a^2_--\textrm{Re}~b^2_-)+\cos\gamma_
   -\sin\gamma_-\textrm{Im}~a_-\textrm{Im}~b_-\right.\nonumber\\
&\left. -1/2\qty(\cos^2\gamma_+-\sin^2\gamma_+)\qty(\textrm{Im}~a^2_+-\textrm{Im}~b^2_+)
 -1/2\qty(\textrm{Re}~a^2_+-\textrm{Re}~ b^2_+)-\cos\gamma_+\sin\gamma_
 +\textrm{Im}~a_+\textrm{Im}~b_+\right]\nonumber\\
&-\sin\alpha\times\nonumber\\
&\left[\cos\gamma_-\qty(\textrm{Im}~a_-\textrm{Re}~b_--\textrm{Im}~b_-\textrm{Re}~a_-)
 +\sin\gamma_-\qty(\textrm{Im}~a_-\textrm{Re}~a_-+\textrm{Im}~b_-\textrm{Re}~b_-)\right.\nonumber\\
&\left.-\cos\gamma_+\qty(\textrm{Im}~a_+\textrm{Re}~b_+-\textrm{Im}~b_+\textrm{Re}~a_+)
-\sin\gamma_+\qty(\textrm{Im}~a_+\textrm{Re}~a_+-\textrm{Im}~b_+\textrm{Re}~b_+)\right],
\end{align}
 \begin{figure}[t]
  \includegraphics[width=0.75\textwidth]{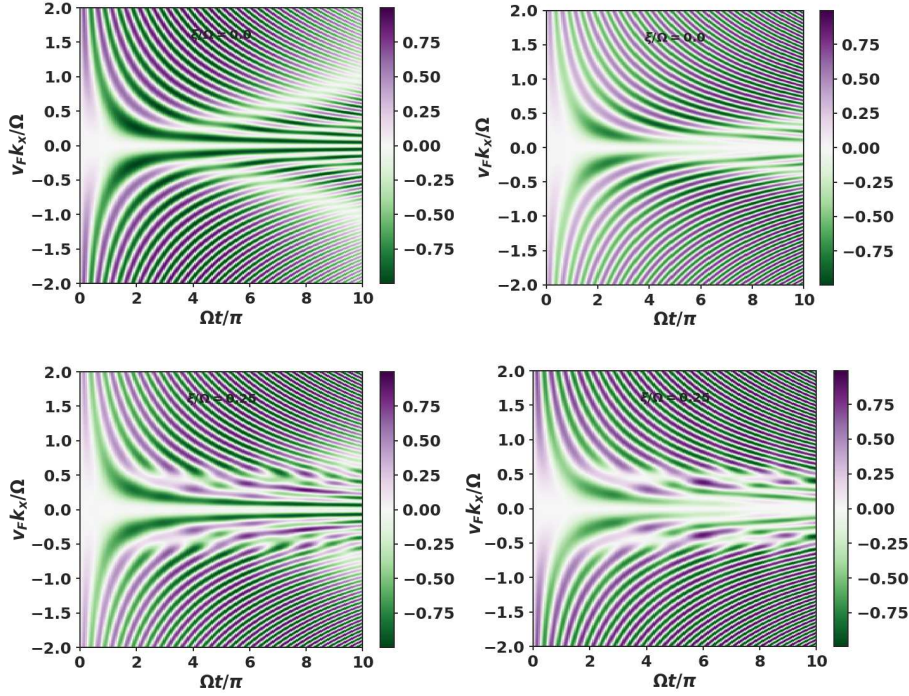}
  \caption{(Color online)
   Effective pseudospin polarization $P_z(t)$  as a function of time and momentumfor the parameters chosen as
   in~\figurename{~\ref{fig:3}}.
   Near the Dirac point ($k=0$) the driving field effectively induces pseudospin inversion at finite driving.}
   \label{fig:4}
 \end{figure}
The results are shown in~\figurename{~\ref{fig:4}} for the same parameters as in fugure \figurename{3}. We
note that contrary to what happens for $S_z(t)$, for
vanishing value of the driving field strength $\xi=0$ (upper panels) show a qualitative behaviour which is not
only dominated by the Rashba spin-orbit interaction $\lambda_R=0.05\Omega$. First we notice that the dynamics
of $P_z(t)$ is much faster than that of $S_z(t)$ since more spin-flipping processes occur within a shorter
time window. These might stem from the fact that even without Rashba spin-orbit interaction, the pseudospin
is a nonconserved quantity since the static Hamiltonian is nondiagonal (even if $\lambda_R=0$) for finite momentum. In addition, in the lower panels (corresponding to finite driving) we notice that there is a
more effective modulation of the pseudospin polarization. This is a natural consequence of the effective role
of perpendicular electric fields acting differently at each sublattice in silicene. Moreover,
from the figures in the lower panels of~\figurename{~\ref{fig:4}} we can infer that driving the system in the
single-valley Dirac cone configuration of~\figurename{~\ref{fig:1}} (lower left panel) and the gapless scenario 
of~\figurename{~\ref{fig:2}} (lower right panel) might be distinguished from each
other by selecting those resonant procesesses happening in the vicinity of the resonances $v_Fk/\Omega=\pm0.5$ where pseudospin-flipping might be accounted for the
avoiding crossings than can be observed around these values of momenta that qualitatively can be understood as an admixture of both pseudospin components with
equal weights. We also notice in the lower left (right) panel that for small momenta, the long term dynamics
$\qty(\Omega t/\pi\ge4)$ the trivial (nontrivial) topological scenario is dominated by vanishing (finite) value
of the pseudospin close to zero momentum.
\section{conclusions}
We have shown that the realization of effective spin-pseudospin exchange requires the interplay of both the spin nonconserving terms associated to 
the rather weak Rashba spin-orbit term as well as the time-dependent field strength. In fact, for the configuration where both of the effective spin sectors have  
an energy spectrum that is gapped at $k=0$, the cooperative effects of the Rashba term and the driving interaction are enhanced as the spin polarization shows additional
flipping processes which require a shorter phase difference among the two spin components. Since this phase difference shifts the maxima and minima of $S_z(t)$,
additional minima appear in its dynamical evolution. Indeed we could expect that the closer in value the quasienergy spectra of the two effective 
spin states are, the probability for spin-flipping processes would in turn be reduced and the additional minima should disappear. We think that these findings
could be interesting for a better understanding of the role of Rashba spin-orbit coupling in silicene and our results show that this previously
disregarded spin interaction in silicene could indeed be accounted for in experimental setups where fine control of the spin degree of freedom would be required. 
For an actual experimental realization, 
we could expect that standard spectroscopic techniques could allow the validation of our results, for instance via time-resolved spin-dependent photoemission 
measurements. We also have found that resonant processes play an important role in pseudospin dynamics and the evolution of these two quantities could afford a 
probe for the dynamical evolution and possible modulation of the topological properties of these new materials\cite{cacho_momentum-resolved_2015} for instance, by time-resolved ultrafast 
optical spectroscopy studies of topological insulators\cite{wang_unraveling_2016} 
or by using time, spin and angle-resolved photoemission of the ultrafast dynamics of topological insulators\cite{sanchez-barriga_subpicosecond_2017} that can be extended to systems where large values of
the Rashba spin-orbit interaction could be achieved in group IV films\cite{zhang_two-dimensional_2017}.
\acknowledgments
This work has been supported by Deutsche Forschungsgemeinschaft via GRK 1570 ``Electronic Properties of Carbon Based Nanostructures''. F. M. and A. L. acknowledge the support of PAPIIT-UNAM through the project IN111317.
\bibliography{biblio}
\end{document}